\documentclass[preprint,12pt]{elsarticle}

\usepackage{amssymb}

\usepackage{graphicx}
\usepackage{microtype}
\usepackage{amssymb}
\usepackage{amsthm}
\usepackage{amsmath}
\usepackage{caption}

\usepackage{booktabs}

\usepackage{hyperref}
    \hypersetup{
        colorlinks   = true,
        citecolor    = blue
    }

\journal{Nuclear Physics B}

\begin{document}
\captionsetup[figure]{labelfont={bf},labelformat={default},labelsep=period,name={Fig.}}
\captionsetup[table]{labelfont={bf},labelformat={default},labelsep=none,singlelinecheck=false,name={Table}}
\captionsetup[table]{skip=2pt}
\begin{frontmatter}



\title{MSW-Transformer: Multi-Scale Shifted Windows Transformer Networks for 12-Lead ECG Classification}


\author[address1]{Renjie Cheng}
\ead{renjie926@foxmail.com}

\author[address1]{Zhemin Zhuang}
\ead{zmzhuang@stu.edu.cn}

\author[address2]{Shuxin Zhuang\corref{cor1}}
\ead{shuxin613@qq.com}

\author[address3]{Lei Xie\corref{cor1}}
\ead{15917917101@163.com}

\cortext[cor1]{Corresponding author}

\author[address4]{Jingfeng Guo}

\address[address1]{Department of Electronic Engineering, Shantou University, No.243, Daxue Road, Tuo Jiang Street, Jinping District, Shantou City, Guangdong, China}
\address[address2]{Department of Biomedical engineering, Sun Yet-sun University, No.66, Guangming District, Gongchang road,  Shenzhen City, Guangdong, China}
\address[address3]{Department of The First Affiliated Hospital of Shantou University, 57 Changping Road, Longhu District, Shantou, Guangdong, China}
\address[address4]{Research Centre, Shantou Institute of Ultrasonic Instruments Co., Ltd., China}

\begin{abstract}
Automatic classification of electrocardiogram (ECG) signals plays a crucial role in the early prevention and diagnosis of cardiovascular diseases. While ECG signals can be used for the diagnosis of various diseases, their pathological characteristics exhibit minimal variations, posing a challenge to automatic classification models. Existing methods primarily utilize convolutional neural networks to extract ECG signal features for classification, which may not fully capture the pathological feature differences of different diseases. Transformer networks have advantages in feature extraction for sequence data, but the complete network is complex and relies on large-scale datasets. To address these challenges, we propose a single-layer Transformer network called Multi-Scale Shifted Windows Transformer Networks (MSW-Transformer), which uses a multi-window sliding attention mechanism at different scales to capture features in different dimensions. The self-attention is restricted to non-overlapping local windows via shifted windows, and different window scales have different receptive fields. A learnable feature fusion method is then proposed to integrate features from different windows to further enhance model performance. Furthermore, we visualize the attention mechanism of the multi-window shifted mechanism to achieve better clinical interpretation in the ECG classification task. The proposed model achieves state-of-the-art performance on five classification tasks of the PTBXL-2020 12-lead ECG dataset, which includes 5 diagnostic superclasses, 23 diagnostic subclasses, 12 rhythm classes, 17 morphology classes, and 44 diagnosis classes, with average macro-F1 scores of 77.85\%, 47.57\%, 66.13\%, 34.60\%, and 34.29\%, and average sample-F1 scores of 81.26\%, 68.27\%, 91.32\%, 50.07\%, and 63.19\%, respectively. The excellent performance of the MSW-Transformer model and its ability to classify different types of diseases demonstrate its potential for assisting heart disease experts in detecting various diseases and obtaining more accurate and comprehensive diagnostic results.
\end{abstract}




\end{frontmatter}


\section{Introduction}
\label{sec:introduction}

According to the estimation of the World Health Organization~\cite{WHO2019}, the number of deaths due to cardiovascular diseases reaches 17.9 million every year, accounting for 32\% of the total global deaths, which is the main cause of human death worldwide. As one of the most commonly used tools for the prevention and diagnosis of cardiovascular diseases, electrocardiogram (ECG) has the advantages of convenience, non-invasiveness and low cost. The electrocardiogram records the electrical activity of the heart. However, the process of analyzing the ECG requires a highly specialized doctor and is labor-intensive. At the same time, doctors are subjective in analyzing ECG, which is easy to cause diagnostic errors~\cite{salerno2003competency}. Therefore, computer-aided diagnosis (CAD) methods are needed to reduce the burden on doctors and reduce subjective errors. In the past, many methods for automatic disease classification based on ECG signals have been developed. The earliest work used traditional ECG classification methods based on signal decomposition. With the rise of machine learning, people have proposed methods that rely on traditional machine learning. However, these methods rely on artificial feature extraction. The selection of features is based on experience and professional knowledge. There are subjectivity and limitations, and the classification accuracy is low. In recent years, with the continuous development of deep learning technology, methods based on deep learning have been widely used in ECG signal classification. Deep learning methods can automatically extract features from original signals, have stronger adaptability and scalability, and can better solve the problems existing in traditional methods, making ECG signal classification more accurate and efficient.

However, traditional deep learning networks still have some limitations. For the processing of long ECG signal sequences, traditional deep learning networks such as Recurrent Neural Network (RNN)~\cite{rumelhart1985learning} and Long Short-Term Memory (LSTM)~\cite{hochreiter1997long} often have the problem of gradient disappearance or gradient explosion~\cite{wulan2020generating}, while Transformer network adopts self-attention mechanism, which can efficiently process sequences of any length and avoid these problems. At the same time, the traditional deep learning network often ignores the dependencies between different positions in ECG signals, while the Transformer network introduces a position coding mechanism, which can better learn the characteristics of different positions in ECG signals, thereby improving the accuracy of feature extraction. In addition, the Transformer model uses a multi-head attention mechanism, which can capture different sequence features at different time steps, further improving the effect of characterizing ECG signal sequences. Due to the aforementioned advantages, Transformer networks have been widely applied in the field of time series classification. Liu et al.~\cite{liu2021gated} proposed Gated Transformer Networks (GTN), With the gating that merges two towers of Transformer which model the channel-wise and step-wise correlations respectively. Better results have been achieved in 13 multivariate time series classifications. Yuan and Lin ~\cite{9252123} proposed a self-supervised pre-trained Transformer network for original optical satellite time series classification and achieved better performance.

In view of the above challenges, this paper proposes a Multi-scale Shifted Windows Transformer Networks (MSW-TN) based on multi-scale sliding window attention mechanism to extract ECG signal features at different scales for classification. In this network, the shifted window restricts self-attention to non-overlapping local windows. This approach effectively extracts information from multiple scales of ECG signals while preserving signal integrity, reducing network complexity and computational cost, and effectively mitigating overfitting risks. The multi-window attention mechanism is similar to the approach used by clinical doctors in diagnosing ECG signals, where relevant features from a local time window of interest are selected for analysis, similar to the multi-window attention mechanism in the model that focuses solely on the relevant features of local time windows. To reduce information loss during multiple window feature extraction and improve model performance, we propose a trainable Multi-Scale Windows feature fusion method (MSW-Feature fusion). Our model also employs attention score visualization techniques that intuitively and clearly present the contribution of different windows to the classification results, enabling the evaluation of the interpretability and accuracy of our model and allowing clinicians and researchers to better understand and interpret the classification results. We evaluate the model on the PTB-XL dataset~\cite{wagner2020ptb}, which is the largest publicly available ECG dataset to date, and compare its performance with that of existing advanced network models, achieving the best results.

Overall, the major contributions in this paper are as follows:

1)We design a new transformer block, named MSW-Transformer Block, based on a multi-scale sliding window attention mechanism, which enables the extraction of multi-dimensional features of ECG signals from different scales. The MSW-Transformer model adopts the design of a single-layer MSW-block, which not only maintains good performance but also further reduces the complexity and computational cost of the model.

2)We propose a trainable feature fusion method, named MSW-Feature fusion, to better leverage the feature information extracted from multiple windows. Specifically, a trainable weight is assigned to each window to adjust the contribution of the features extracted from that window. By fusing the features based on the learned weights, the proposed method can enhance classification accuracy.

3)We investigated the attention visualization of multi-scale window attention mechanism models of different scales in ECG classification tasks to achieve better clinical interpretability.

4)We conducted the evaluation of all classification tasks on the publicly available PTB-XL dataset for the first time. The experimental results, along with the comparison of performance against state-of-the-art methods, validate the effectiveness of the proposed MSW-TN (Multi-Scale Shifted Windows Transformer Networks).

The rest of the paper is organized as follows. Section 2 introduces the related work on ECG signal classification. Section 3 presents our proposed method. Section 4 describes the dataset used in our experiments. Sections 5 and 6 discuss our experiments and results. Section 7 discuss some details of the paper. Finally, Section 8 summarizes the paper.

\section{Related Works}
\label{sec:Related Works}

\subsection{Overview of ECG Signal Classification Methods}

The literature on ECG signal classification can be divided into three stages. In the first stage, traditional ECG signal classification methods based on signal decomposition were proposed, which classified normal and arrhythmic signals using discrete cosine transform (DCT) and discrete wavelet transform (DWT)~\cite{2012Classification}. Martis et al.~\cite{martis2013ecg} used principal component analysis (PCA) and independent component analysis (ICA) to classify ECG signals. In the second stage, machine learning-based methods were proposed. Naike et al.~\cite{naik2016classification} used the multi-model decision learning (MDL) algorithm to achieve better sensitivity in ECG classification as normal and abnormal. Goovaerts et al.~\cite{goovaerts2018machine} used variational mode decomposition (VMD) to segment the QRS complex in ECG signals and input calculated features to classifiers such as Support Vector Machine (SVM) and K-Nearest Neighbors (KNN). Kiranyaz et al.~\cite{kiranyaz2015real} used a Convolutional Neural Network (CNN) to classify cardiac arrhythmias in two-dimensional ECG signals. In the third stage, deep learning-based methods have shown outstanding performance in ECG signal classification research, achieving higher accuracy and efficiency than expert manual classification. Rajpurkar et al.~\cite{rajpurkar2017cardiologist} trained a 34-layer convolutional neural network that mapped a sequence of ECG samples to a sequence of rhythm classes. Liu et al.~\cite{liu2018multiple} proposed a multi-feature branch CNN, where each lead in the 12-lead signal corresponds to each branch. Chen et al.~\cite{chen2020automated} combined CNN and LSTM to classify six types of ECG signals. Ahmed et al.~\cite{mostayed2018classification} proposed an RNN and two bidirectional LSTMs feature classifier that can train ECG signals of any duration. Patrick et al.~\cite{schwab2017beat} used RNN with attention mechanism for single-channel ECG signal classification. Mousavi et al.~\cite{mousavi2019ecgnet} proposed an automatic heartbeat classification method based on a deep convolutional neural network and a sequence-to-sequence model. Śmigieli et al.~\cite{smigiel2021ecg} proposed a convolutional network with entropy features for ECG classification, and added extracted QRS complex features for ECG classification~\cite{smigiel2021deep}. Pałczyński et al.~\cite{palczynski2022study} proposed using few-shot learning for ECG classification. Feyisa et al.~\cite{feyisa2022lightweight} designed a multi-receptive field CNN architecture for ECG classification. Reddy et al.~\cite{reddy2021imle} proposed an ECG classification model called IMLE-Net, which utilizes multi-channel information available in standard 12-lead ECG recordings and learns patterns at beat, rhythm, and channel levels. Murugesan et al.~\cite{murugesan2018ecgnet} presented a deep neural network called ECGNet for the classification of cardiac arrhythmias, which is composed of one-dimensional convolutions and bidirectional LSTM layers.

\subsection{Overview of Time Series Classification based on Transformer Networks}

Transformer networks were first proposed in the field of natural language processing (NLP)~\cite{vaswani2017attention} and applied to machine translation tasks with excellent results. With the successful application of Transformer networks in NLP, their applications in other fields have also gradually attracted the attention of researchers. For example, Vision-Transformer network (Vi-TN)~\cite{dosovitskiy2020image} and Swin-Transformer network (Swin-TN)~\cite{liu2021swin} are proposed to provide more efficient and accurate models in the field of computer vision (CV). Transformer network has also been widely applied in the field of time series classification and has achieved outstanding performance. Liu {et al.}~\cite{liu2021gated} proposed Gated Transformer Networks, which exhibited improved performance on 13 multivariate time series classification tasks. Yuan and Lin~\cite{9252123} introduced a self-supervised pre-training Transformer network designed for the classification of time series data in satellite optical imagery. Additionally, Yuan {et al.}~\cite{yuan2022sits} introduced a pre-training model called SITS-Former, which is used for Satellite Image time series classification. Deep learning-based ECG classification, which comes at the cost of increasing the complexity of the network architecture, has achieved success, especially in state-of-the-art models where the number of layers and neurons per layer is increasing. Specifically, in Transformer-based models, there are more layers, higher model complexity, and greater computational cost, which can lead to overfitting on small-scale datasets.

\subsection{Overview of Visualization Techniques}

The opaque nature of deep learning models poses a challenge to the clinical interpretation of ECG classification, as these models often lack transparency in explaining the underlying basis for their classification decisions, which may impede doctors' comprehension and application of classification outcomes. An interpretable model with high accuracy is crucial for the clinical application of computer-aided ECG diagnosis. Currently, many researchers have made considerable efforts to improve the interpretability of deep learning models in computer vision tasks. However, interpreting ECG classification models has been a limited yet notable area of research. Vijayarangan \emph{et al.}~\cite{vijayarangan2020interpreting} proposed gradient-weighted class activation maps to visualize their single-channel ECG signal classification model. Reddy \emph{et al.}~\cite{naik2016classification} visualized attention scores of ECG classification model results to achieve better clinical interpretation. For the visualization of Transformer models, Kobayashi \emph{et al.}~\cite{kobayashi2020attention} utilized multiple colors to represent the attention matrices of different layers and displayed their relationships graphically. Strobelt \emph{et al.}~\cite{strobelt2018debugging} attempted to visualize the attention matrix over time, allowing the user to observe the specific locations of the model's focus when processing sequences. Chefer \emph{et al.}~\cite{chefer2021transformer} proposed a visualization method for Transformer networks that integrates attention and correlation scores into multiple attention modules and includes normalization terms for non-parametric layers.

\section{PROPOSED METHODOLOGY}

\begin{figure}[ht]
\centering 
\centerline{\includegraphics[width=1\linewidth]{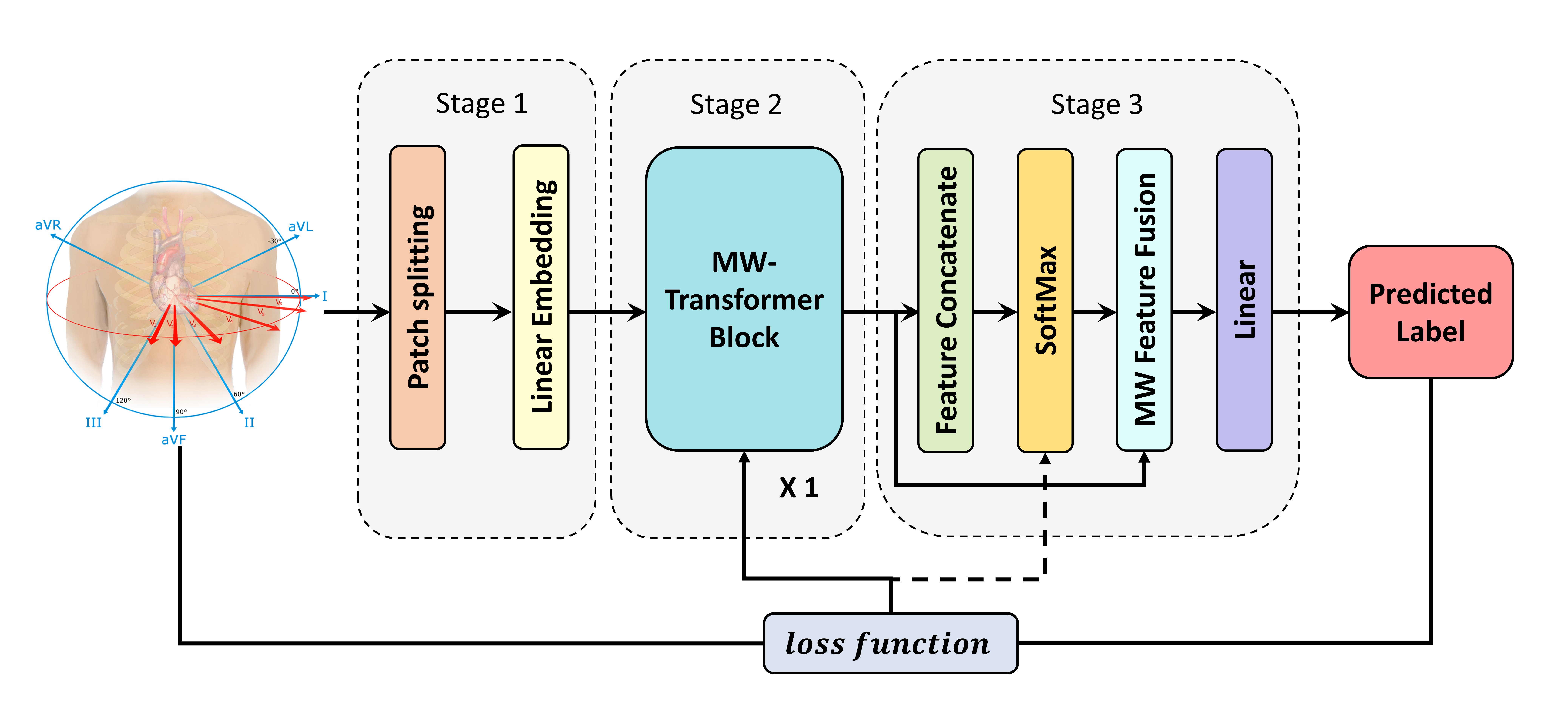}}
\caption{Model architecture of the Multi-Scale Shifted Windows Transformer Networks.}
\label{fig1}
\end{figure}

\subsection{Overview of Time Series Classification based on Transformer Networks}

The overview of the overall architecture of the MSW-Transformer network is presented in Fig.~\ref{fig1}, which is primarily composed of three steps:

The first stage of the MSW-Transformer network consists of signal preprocessing, as shown in Stage 1 of Fig.~\ref{fig1}. This stage includes two sub-modules, namely the patch splitting module (PSM) and the linear embedding module (LEM). Firstly, the 12-lead ECG signal is processed using the PSM, which segments the signal into non-overlapping patches. This approach extracts local features, treating each patch as a token, thereby improving classification accuracy and reducing model computational cost. Next, the LEM is applied to extract important features from the raw data, reducing the influence of noise and redundant information, and mapping the original feature dimension to a dimension that MSW-Transformer block can accept. The processed signal is then input into Stage 2.

The second stage of the MSW-Transformer network involves feature extraction, using the designed MSW-Transformer block, as shown in Stage 2 of Fig.~\ref{fig1}. After inputting the signal into the MSW-Transformer block, this stage extracts features from the signal. We employ multi-head self-attention (MSA) mechanism within three shifting windows of different scales to extract multidimensional local features of electrocardiogram signals, including spatial and temporal dimensions as well as positional information. This approach significantly reduces model complexity and computational cost, while improving inference speed and classification accuracy. Features are extracted from the tokens inputted in Stage 1, and the feature information extracted by windows of different scales represents features with different receptive fields.

The third stage involves feature fusion using the proposed MSW-Feature Fusion approach, a trainable multi-window feature fusion approach, as depicted in Stage 3 of Fig.~\ref{fig1}. In Stage 2, features with different receptive fields were extracted using windows of three different scales. However, how to effectively fuse these features is a critical issue. To make optimal use of the features extracted in Stage 2, a trainable weight is assigned to each window in Stage 3 to control the degree of feature fusion. Specifically, the features obtained from each window are fused according to their weights, which reflect their importance for classification. This approach effectively utilizes the multi-window extracted feature information, thereby enhancing classification accuracy.

\subsection{Signal Preprocessing}

In order to improve the performance and efficiency of MSW-TN, two preprocessing steps need to be performed on the ECG signals.

We partition the 12-lead ECG signal into non-overlapping patches through a PSM, where each patch is regarded as a token with its feature being the concatenation of the raw signal values from the 12 leads. This approach associates the original features between leads for the raw ECG signal. Since the feature of each token can more accurately describe local signal changes and characteristics, this approach facilitates the subsequent extraction of local features. The model can extract features from each token more accurately, thereby improving the overall accuracy of the classification results. Furthermore, since the computation cost of extracting features from each token is relatively small compared to processing the entire ECG signal, feature extraction on tokens can also reduce the computational cost of the model. For the specific implementation, we use the ECG signals from the PTB-XL dataset, with $n=12$ channels and a length of $L=1000$ for each channel. The model uses patch size of $1\times5$, resulting in a feature dimension of $1\times5\times12=60$ for each patch. Subsequently, a linear projection is applied to project the raw features into a linear dimension of $C=512$, which matches the receptive dimension of the MSW-transformer block. After the linear embedding, these tokens are then inputted into the MSW-transformer block.

\subsection{MSW-Transformer Block}

\begin{figure}[htbp]
\centering 
\centerline{\includegraphics[width=.6\linewidth]{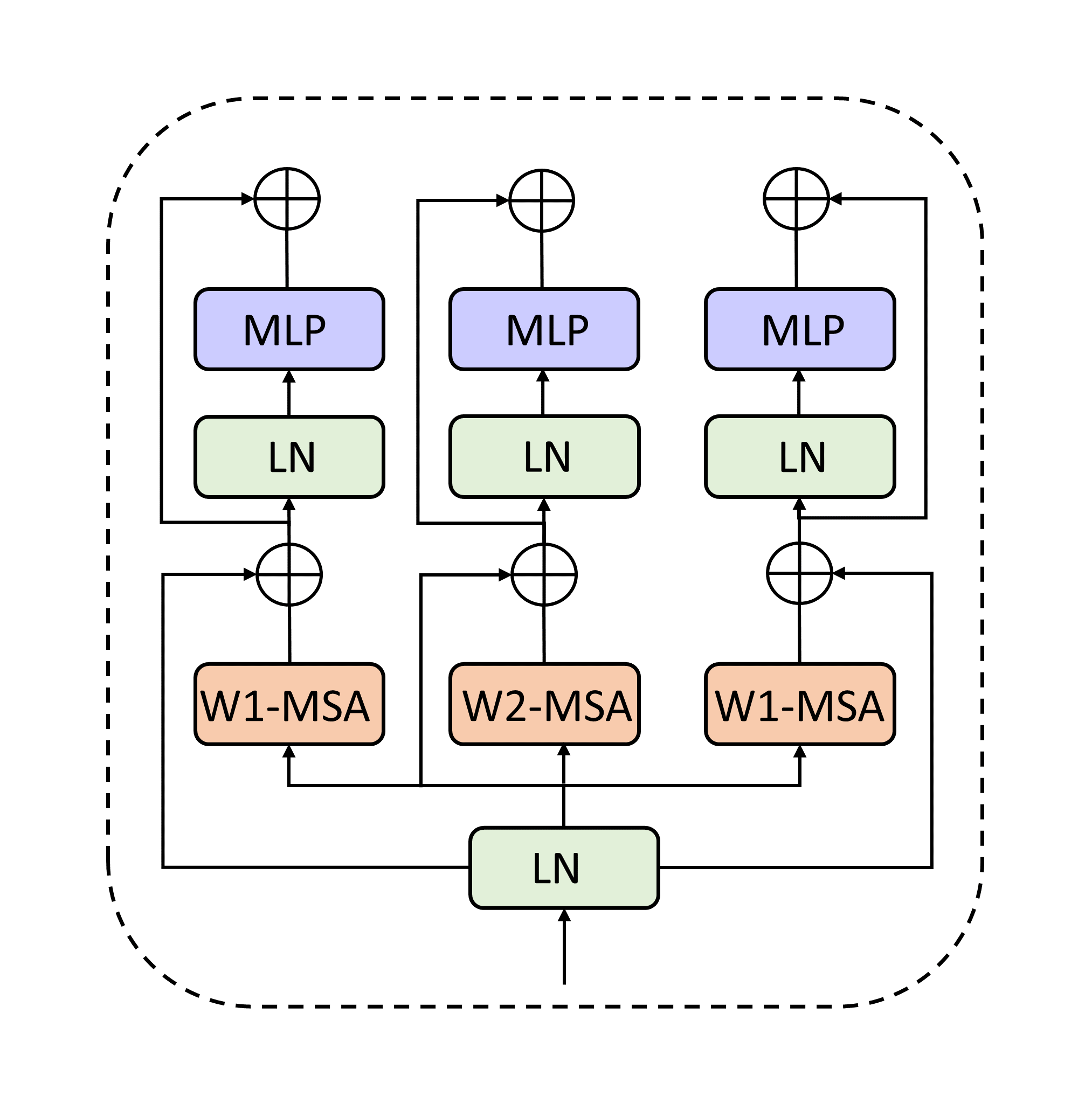}}
\caption{Network architecture of MSW-Transformer block.}
\label{fig2}
\end{figure}

The network architecture of the MSW-Transformer block is illustrated in Fig.~\ref{fig2}, which comprises three windows of different sizes, $1\times M_i(i\in1,2,3)$, where $M_i (i\in1,2,3)$ represents the varying lengths of the windows. Each window independently computes the MSA mechanism, followed by a 2-layer Multilayer Perceptron (MLP) with GELU non-linear activation layers. Before each MSA module and each MLP, a LayerNorm (LN) layer is applied, and residual connections are applied after each module. Our network uses only one layer of the MSW-Transformer block, which differs from the conventional Transformer network. This notably reduces the parameters and computational cost of the network. For the ECG signal classification task, a single layer of the MSW-Transformer block can achieve effective feature extraction results.

\begin{figure}[htbp]
\centering 
\centerline{\includegraphics[width=1\linewidth]{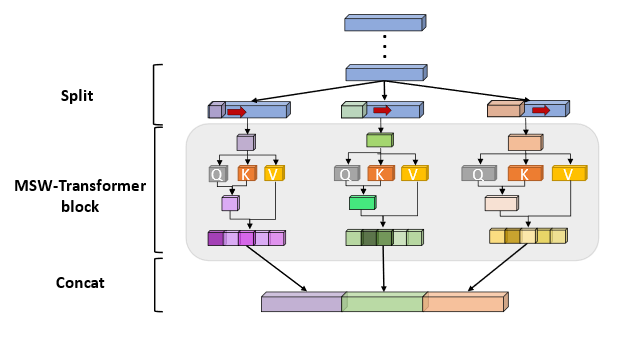}}
\caption{The Three Steps of ECG Signal Feature Extraction.}
\label{fig3}
\end{figure}

Feature extraction is divided into three stages, as shown in Fig.~\ref{fig3}. In the first stage, the splitting stage, the ECG signal output from the LM is split into separate time blocks based on three window lengths. In the second stage, features are extracted using MSW-transformer block, and the features obtained from each window are connected to generate three feature vectors of different lengths. In the third stage, the linear transformation stage, the three feature vectors obtained from the second stage are passed through a linear layer and then enter stage 3.

\subsubsection{Attention Mechanism}

The self-attention algorithm used in the MSW-Transformer block follows the approach proposed in Swin-TN~\cite{liu2021swin}, and its effectiveness has been demonstrated. Specifically, it includes the relative positional bias $B\in\mathbb{R}^M$ when computing self-attention:

\begin{equation}
\mathrm{Attention}(Q,K,V)=\mathrm{SoftMax}(\frac{QK^T}{\sqrt d}+B)V 
\label{eq1}
\end{equation}

In Eq.~\eqref{eq1}, $Q,K,V\in \mathbb{R}^{M\times d}$ are the $query$, $key$ and $value$ matrices, where $d$  is the dimensionality of the $query$ and $key$, and $M$ is the number of patches. Since the relative positional offsets along the horizontal axis fall within the range of $[-M+1,\ M-1]$, we parameterize a small bias matrix $\hat{B}\in \mathbb{R}^{2M-1}$, and the values of $B$ belong to $\hat{B}$. Incorporating relative positional bias into the attention mechanism can facilitate the more effective capture of local features within a sequence, thereby promoting the modeling of sequence information by the model. It can also avoid elements at different positions in the sequence having the same impact on attention weights, thereby improving the expressiveness and performance of the model.

\subsubsection{Computational Complexity}

\begin{equation}
\mathrm{Attention}(Q,K,V)=\text{SoftMax}\left( \frac{QK^T}{\sqrt{d}} \right) V
\label{eq2}
\end{equation}

The global self-attention mechanism of the conventional Transformer architecture, as computed in Eq.~\eqref{eq2}, operates on each token with respect to all other tokens, resulting in a quadratic computational complexity in relation to the number of tokens. Hence, it is unsuitable for prediction or classification tasks that involve a large number of tokens. To compare the computational complexity of MSA and self-attention mechanisms based on the MSW-Transformer block (MSW-SA), we conducted a complexity analysis.

For a multidimensional signal with length  $1\times L$ and $C$ channels:

To compute the complexity of the global attention mechanism in the standard Transformer model, the first step is to generate three feature vectors, namely, $Q$, $K$, and $V$, where $W^Q$, $W^K$, and $W^V$ are weight matrices and the dimension of $W$ is $\left( C,\,\,C \right) $.

\begin{equation}
Q=L\times W^Q\,\,\,\,K=L\times W^K\,\,\,\,V=L\times W^V
\label{eq3}
\end{equation}

Computing the complexity of Eq.~\eqref{eq3} is as follows.

\begin{equation}
\Omega _1=3LC^2
\label{eq4}
\end{equation}

The second step is to compute the complexity of $QK^T$ in Eq.~\eqref{eq2}. The dimensions of $Q$, $K$ and $V$ are $\left( L,\,\,C \right) $, thus the computational complexity of this step is:

\begin{equation}
\Omega _3=L^2C
\label{eq5}
\end{equation}

The third step is to compute $Z$ by multiplying $QK^T$ with $V$ after applying the softmax function according to Eq.~\eqref{eq2}. The dimension of $QK^T$ is $\left( L,\,\,L \right) $, thus the computational complexity of this step is:

\begin{equation}
\Omega _3=L^2C
\label{eq6}
\end{equation}

The fourth step is to compute the final output by multiplying $Z$ with $W^Z$. The computational complexity of this step is:

\begin{equation}
\Omega _4=LC^2
\label{eq7}
\end{equation}

Therefore, the computational complexity of the MSA in the standard Transformer model is:

\begin{equation}
\Omega \left( \text{MSA} \right) =4LC^2+2L^2C
\label{eq8}
\end{equation}

The computational complexity of the MSW-Transformer block is based on the length of the signal, the scale of the windows $\left( 1\times M_i,i\in 1,2,3 \right) $, and the number of channels $C$. To compute the MSA within the window, the complexity for $Q$, $K$ and $V$ remains unchanged. In the second step, $QK^T$ needs to be computed for $\frac{L}{M}$ windows, each with a complexity of $M^2C$. In the third step, the complexity of multiplying with $V$ is also $M^2C$. Therefore, the computational complexity of MSW-Self Attention can be expressed as follows:

\begin{equation}
\Omega \left( \text{MSW}-\text{SA} \right) =4LC^2+2LC\sum_{i=1}^3{M}_i
\label{eq9}
\end{equation}

By comparing Eq.~\eqref{eq8} and Eq.~\eqref{eq9},  it is evident that the computational complexity of MSW-SA is significantly lower than that of MSA. This difference is directly related to the length of the signal $L$. As the signal length increases, the benefits of using a multi-window attention mechanism become more prominent. The use of a global attention mechanism for long signals is computationally expensive and presents challenges for efficient processing. Additionally, applying MSW-Transformer Block in a gradual increase in model computation, effectively managing the model's computational burden even when dealing with long time series.

\begin{figure}[htbp]
\centering 
\centerline{\includegraphics[width=.9\linewidth]{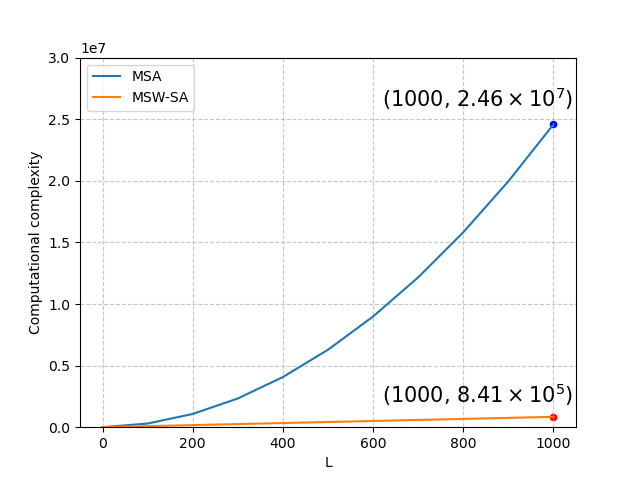}}
\caption{Comparison of the Computational Complexity of MSA and MSW-SA for PTB-XL Dataset, with ECG Signal Length of $L=1000$,  $C=12$, and Window Scales of $M=(1,5)(1,10)(1,20)$.}
\label{fig4}
\end{figure}

As illustrated in Fig.~\ref{fig4}, we have conducted a comparative analysis of the computational complexity between MSA and MSW-SA. It can be observed that the computational complexity of MSA exhibits an exponential growth with the increase of the signal length $L$, whereas the computational complexity of MSW-SA increases at a relatively slower rate, rendering it easier to handle. This finding underscores the superior performance of our proposed MSW-TN in processing long time series, which is particularly beneficial for biomedical signal classification applications that widely employ machine learning, deep learning, and other related technologies.

\subsubsection{Satisfying the Conditions for Window Scales}

For each channel, the signal length was $L$, and patches of size $1\times P$ were used, where each patch had a length of $\frac{L}{P}$. We utilized three window scales, each consisting of $1\times M_i(i\in1,2,3)$ patches. The window scales must satisfy the following conditions:

\begin{equation}
M_i\left( i\in 1,2,3 \right) \mid \frac{L}{P}
\label{eq10}
\end{equation}

\subsection{MSW-Feature Fusion}

In Stage 2, we utilized MSW-transformer block to extract ECG signal features. Effective fusion of these features is crucial, as it directly impacts classification accuracy. To address this, we introduced MSW-Feature fusion, a trainable multi-window feature fusion method that is well-suited for feature fusion in the MSW-Transformer block.

\begin{figure}[ht]
\centering 
\centerline{\includegraphics[width=0.9\linewidth]{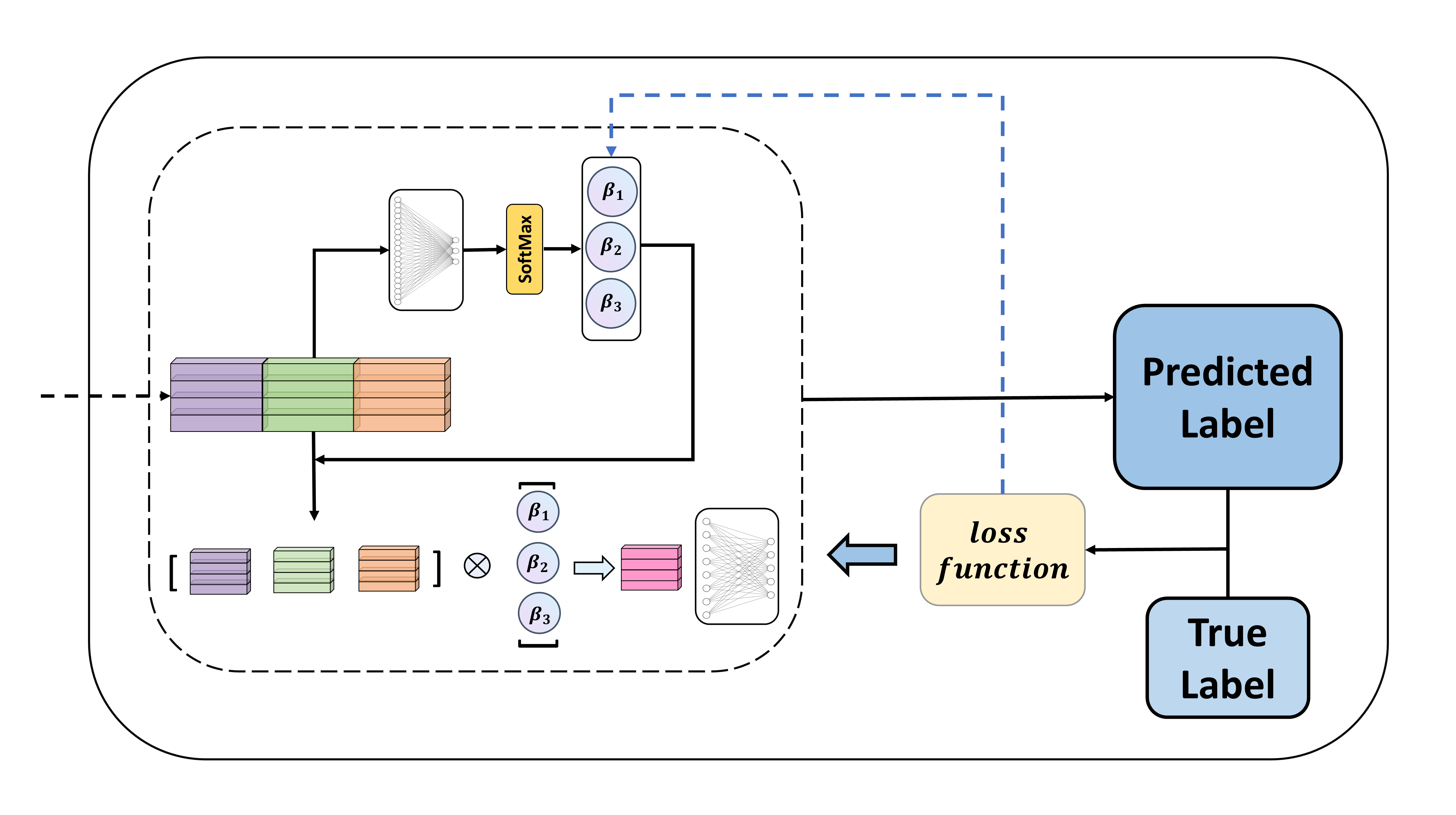}}
\caption{Network architecture of MSW-Feature Fusion.}
\label{fig5}
\end{figure}

As illustrated in Fig.~\ref{fig5}, we proposed a network to train the weights of each window feature. Initially, we concatenated the features extracted from the three windows, $\alpha_1$, $\alpha_2$, and $\alpha_3$, into a vector using a Concatenation operation.

\begin{equation}
\alpha =\text{Concat}\left( \alpha _1,\alpha _2,\alpha _3 \right)
\label{eq11}
\end{equation}

Subsequently, the weights $\beta_1$, $\beta_2$, and $\beta_3$ for each window are obtained via a fully connected layer and the Softmax activation function.

\begin{equation}
\,\,\beta _1,\beta _2,\beta _3=\text{SoftMax}\left( \alpha \right)   
\label{eq12}
\end{equation}

Finally, the features are fused by performing an addition operation and then connected to a sigmoid layer for classification.

\begin{equation}
y=\text{Add}\left( \alpha _1\cdot \beta _1,\alpha _2\cdot \beta _2,\alpha _3\cdot \beta _3 \right) 
\label{qe13}
\end{equation}

The loss in the entire classification model is calculated by predicting the difference between the predicted and actual classification labels. The values of the feature weights, namely $\alpha_1$, $\alpha_2$, and $\alpha_3$ extracted from each window are crucial in determining the classification results and also influence the magnitude of the loss. These parameters are updated continuously during training to minimize the loss. Thus, the network trains iteratively to adjust the feature weights and capture local signal characteristics more effectively, resulting in improved classification accuracy and better results. We conducted experiments to verify the effectiveness of MSW-Feature Fusion, and the experimental results are presented in Fig.~\ref{fig11}.

\section{DATASET OVERVIEW}

\begin{figure}[htbp]
\centering 
\centerline{\includegraphics[width=.6\linewidth]{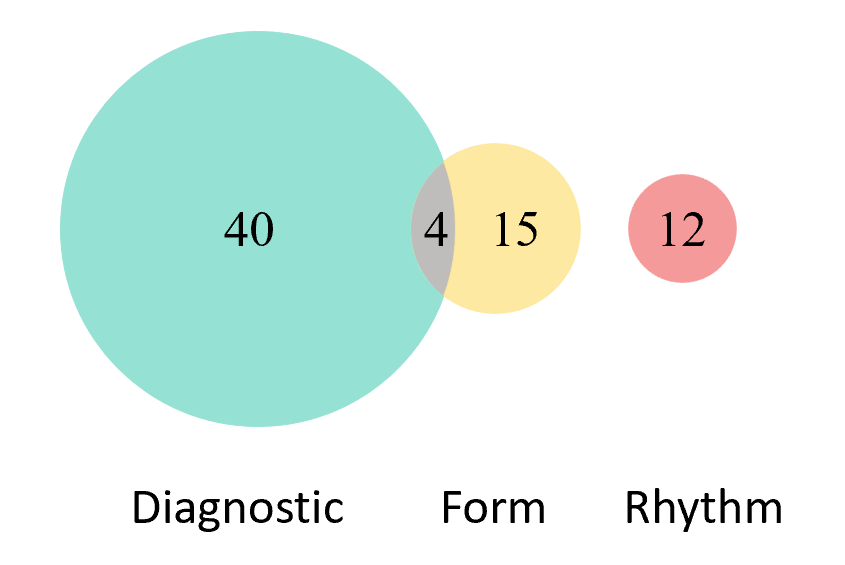}}
\caption{The Venn diagram illustrates the classification of the given SCP ECG statements into the three categories of diagnostic, form, and rhythm.}
\label{fig6}
\end{figure}

The proposed method was assessed on the publicly available PTB-XL~\cite{wagner2020ptb} ECG dataset, which is elaborated upon below. The ECG waveforms in this dataset were annotated by two certified cardiologists, and each ECG record was assigned a label from a set of 71 different labels that adhere to the Standard Communication Protocol for Computer-Assisted Electrocardiography (SCP-ECG) standard. The dataset encompasses a broad spectrum of pathologies with multiple co-occurring diseases, comprising 21,837 10-second, 12-lead ECG signals from 18,885 patients, of which 52\% are male and 48\% are female. The dataset is provided in versions with sampling rates of 500Hz and 100Hz, with our study utilizing the 100Hz version. The dataset is partitioned into three non-exclusive categories, as illustrated in Fig.~\ref{fig6}: diagnostic (e.g., "anterior myocardial infarction"), form (e.g., "abnormal QRS complex"), and rhythm (e.g., "atrial fibrillation"), with a total of 71 different diagnoses, which are further subdivided into 44 diagnostic classes, 19 form classes (four of which are also employed as diagnostic classes), and 12 rhythm classes. The dataset offers a wealth of pathological conditions and diverse classification labels, rendering it suitable for research and applications in ECG signal processing and classification.

\begin{figure}[h!]
\centering 
\centerline{\includegraphics[width=0.8\linewidth]{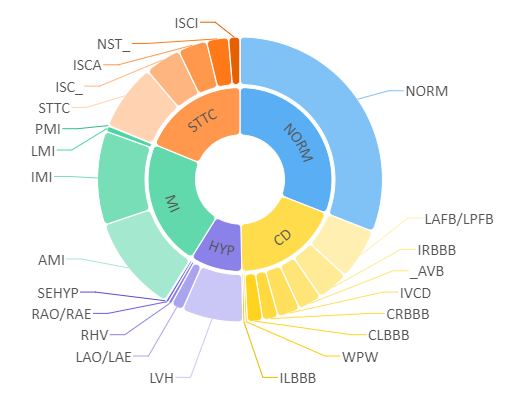}}
\caption{Graphical Summary of Diagnostic Superclasses and Subclasses in the PTB-XL Dataset.}
\label{fig7}
\end{figure}

The diagnostic classes is further categorized into 5 superclasses and 24 subclasses, as illustrated in Fig.~\ref{fig7}. The superclasses comprise Myocardial Infarction (MI), Conduction Disturbance (CD), and other categories, while the subclasses include AMI (Anterior Myocardial Infarction) for MI and IRBBB (Incomplete Right Bundle Branch Block) for CD, among others. Each ECG signal is assigned to one of the ten folds, with the first eight folds used for training the model, the ninth fold used as the validation set, and the tenth fold used as the test set. Prior to training, all training and test data are standardized.

\section{EXPERIMENTAL DETAILS}
\subsection{Classification Task}

We have chosen samples of various categories from the PTB-XL dataset to assess the performance of our model. Our classification tasks are all multi-label classification, comprising five categories as follows:

1)We employed 5 superclass labels in diagnostic categories, as illustrated in Fig.~\ref{fig6}, and categorized them into normal (NORM), conduction disturbance (CD), myocardial infarction (MI), hypertrophy (HYP), and ST/T change (STTC).

2)We employed 23 subclass labels in diagnostic categories, including diseases such as anterior myocardial infarction (AMI), incomplete right bundle branch block (IRBBB), and non-specific ST changes (NST), as shown in Fig.~\ref{fig6}, which belong to the sub-labels of task 1.

3)We employed 12 rhythm class labels, such as sinus rhythm (SR), atrial fibrillation (AFIB), and sinus tachycardia (STACH).

4)We employed 19 form class labels, including non-diagnostic T abnormalities (NDT), ventricular premature complexes (PVC), and high QRS voltage (HVOLT).

5)We employed all diagnostic statements, totaling 44 class labels, including diseases such as anterior myocardial infarction (AMI), sinus rhythm (SR), and non-diagnostic T abnormalities (NDT).

\subsection{Evaluation Metrics}

We will evaluate the proposed model by comparing it with state-of-the-art methods for 12-lead ECG classification based on deep learning, including Resnet 101~\cite{he2016deep}, ECGNet~\cite{murugesan2018ecgnet}, and IMLENet~\cite{reddy2021imle}, as well as the state-of-the-art time series classification method based on the Transformer model, Gated Transformer Networks~\cite{liu2021gated}. Additionally, we will compare the performance of the proposed model with existing models on the same classification task. We will employ standard metrics for classification tasks, including Area under Receiver Operating Characteristics (ROC-AUC), Accuracy, and Average Macro-F1 score for evaluation. Furthermore, we will use an additional metric, Average Samples-F1 score, which essentially evaluates the classification ability of each sample and reflects the model's performance on different samples more comprehensively. This metric is particularly suitable for addressing the issue of data imbalance in the PTB-XL dataset, as it considers the imbalance of sample classes and provides a more comprehensive evaluation of the model's classification performance~\cite{sokolova2009systematic}. Accuracy is defined as the ratio of correct predictions to total predictions, precision is defined as the ratio of true positive (TP) to the sum of true positive and false positive (FP), and recall is defined as the ratio of TP to the sum of TP and false negative (FN).

\begin{equation}
Accuracy=\,\,\frac{TP+TN}{P+N}
\end{equation}
\begin{equation}
Precision=\,\,\frac{TP}{TP+FP}
\end{equation}
\begin{equation}
Recall=\,\,\frac{TP}{TP+FN}\,\,
\end{equation}
\begin{equation}
Macro-F1\ score=\frac{2\times Precision\times Recall}{Precision+Recall}\,\,
\end{equation}
\begin{equation}
Samples-F1\ score\ =\ \frac{\sum{_{i=1}^{n}}\frac{2\times Precision_i\times Recall_i}{Precision_i+Recall_i}}{n}
\end{equation}

\subsection{Interpretable Design}

To assess the interpretability of our model, we will visualize attention scores and compare them with the clinical criteria used by cardiac experts to identify Inferior Myocardial Infarction (IMI). In clinical practice, ST-segment elevation in leads II, III, and aVF of ECG signals has been demonstrated as one of the criteria for diagnosing IMI~\cite{morris2002abc, edhouse2002acute}. Therefore, we aim to demonstrate the feasibility and effectiveness of our model by visualizing the attention scores of the ST segment in leads II, III, and aVF of IMI cases.

\subsection{Implementation Details}

The model was trained for a maximum of 50 epochs using a batch size of 16 or 32 for different tasks. Cross-Entropy loss function was employed with the Adam optimizer~\cite{2014Adam}, and the learning rate was set to 0.0001, which was reduced by a factor of 10 every 10 epochs to observe improvements in results. To prevent overfitting, the dropout rate for the attention mechanism of the shifting window was set to 0.2. The model was implemented using the Pytorch framework and was trained on an Nvidia RTX 3090 24GB GPU workstation.

\section{EXPERIMENTAL RESULTS}

\subsection{Model Complexity}

\begin{figure}[h]
\centering 
\centerline{\includegraphics[width=1\linewidth]{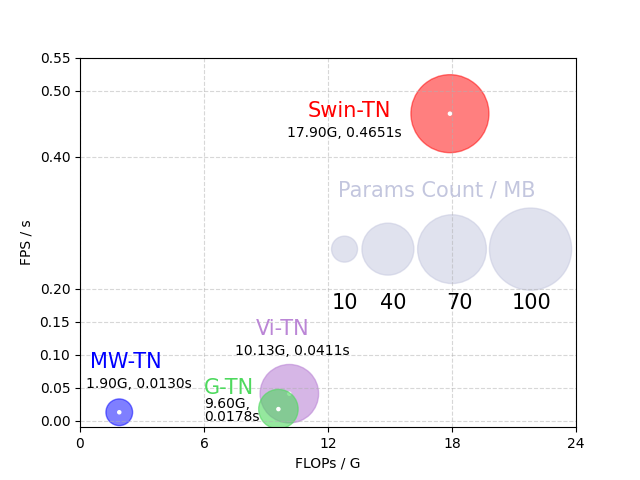}}
\caption{Comparison Results of Swin-TN, Vi-TN, G-TN, and MSW-TN in Model Complexity, Theoretical Computational Complexity, and Inference Speed.}
\label{fig8}
\end{figure}

To demonstrate the superiority of our proposed model, as illustrated in Fig.~\ref{fig8}, we conducted a comparative analysis against a range of advanced Transformer models, including Swin-TN~\cite{liu2021swin}, Vi-TN~\cite{dosovitskiy2020image}, and G-TN~\cite{liu2021gated}, in three key performance metrics: model size (parameter), theoretical computational complexity (FLOPs), and inference speed (FPS). The evaluation was performed on the task of classifying five diagnostic superclass labels, utilizing the same GPU workstation. Compared to Swin-TN, MSW-TN has a parameter size that is 0.12 times smaller, a computational complexity that is 0.11 times lower, and an inference speed that is 35.78 times faster. Compared to Vi-TN, MSW-TN has a parameter size that is 0.20 times smaller, a computational complexity that is 0.18 times lower, and an inference speed that is 3.16 times faster. Compared to G-TN, MSW-TN has a parameter size that is 0.46 times smaller, a computational complexity that is 0.20 times lower, and an inference speed that is 7 times faster. During testing using the aforementioned Transformer models, all models except for G-TN exhibited overfitting during the testing process.

\subsection{Classification Results}

Tables~\ref{tab1}-\ref{tab5} present the results of our proposed model on five distinct classification tasks. Our study marks the first complete application of a Transformer-based model for ECG signal classification. Notably, MSW-TN outperforms other models, achieving the highest performance in both Macro-F1 score and Samples-F1 score.

\renewcommand{\arraystretch}{1.2}

\begin{table}[!ht]
\caption{\newline Model evaluation for classification of diagnostic superclass labels on the PTB-XL dataset.}
\setlength{\tabcolsep}{0.0008mm}
\begin{tabular}{lccccc}
\hline
 &
  Accuracy(\%) &
  \begin{tabular}[c]{@{}c@{}}ROC-AUC \\ Macro(\%)\end{tabular} &
  \begin{tabular}[c]{@{}c@{}}ROC-AUC \\ Samples(\%)\end{tabular} &
  \begin{tabular}[c]{@{}c@{}}Macro-F1 \\ score(\%)\end{tabular} &
  \begin{tabular}[c]{@{}c@{}}Samples-F1 \\ score(\%)\end{tabular} \\ \hline
Smigiel et al.~\cite{smigiel2021ecg}    & 76.30          & 90.30          & -               & 68.30          & -               \\
Pałczynski et al.~\cite{palczynski2022study} & 79.00          & \textbf{93.60} & -               & 70.60          & -               \\
Feyisa et al.~\cite{feyisa2022lightweight}    & 89.70          & 93.00          & -               & 72.00          & -               \\
IMLENet~\cite{reddy2021imle}           & 88.85          & 92.16          & -               & -               & -               \\
GTN              & 81.30          & 86.77          & 88.91          & 68.87          & 72.04          \\
IMLENet                 & 88.85          & 92.16          & 91.75          & 73.91          & 76.37          \\
ECGNet                  & 87.50          & 90.55          & 91.29          & 75.31          & 78.15          \\
Resnet101              & 87.87          & 91.19          & \textbf{92.10} & 76.19          & 78.47          \\
\textbf{MSW-TN}         &\textbf{89.80} & 92.33          & 91.98          &\textbf{77.85} & \textbf{81.26} \\ \hline
\end{tabular}
\label{tab1}
\end{table}

The results for Task 1, which involves classifying the superclasses labels, are shown in Table~\ref{tab1}. The first four models, Smigiel et al.~\cite{smigiel2021ecg}, Pałczynski et al.~\cite{palczynski2022study}, Feyisa et al.~\cite{feyisa2022lightweight}, and IMLENet~\cite{reddy2021imle}, cited results directly from the literature, while the results for the latter four models, GTN, IMLENet, ECGNet, and Resnet101, were obtained through our own experiments. Our proposed model achieved the best performance in terms of Accuracy, Macro-F1 score, and Samples-F1 score, with scores of 89.80\%, 77.85\%, and 81.26\%, respectively. Specifically, clinical diagnosis of superclasses such as Conduction Disorders (CD) and Hypertrophy (HYP) require diagnostic criteria such as QRS wave width and amplitude abnormalities~\cite{surawicz2009aha}. Our small-scale window attention mechanism is well-suited for extracting these changing features of QRS waves. In clinical diagnosis of Myocardial Infarction (MI) and ST/T segment (STTC), the appearance of ST segment elevation or depression is one of the diagnostic criteria~\cite{rautaharju2009aha}. Our large-scale window attention mechanism is better equipped for extracting these changing features. Overall, our proposed MSW-Transformer can effectively extract features from electrocardiogram signals and achieve better feature fusion, resulting in superior classification results.

\begin{table}[!ht]
\caption{\newline Model evaluation for classification of diagnostic subclass labels on the PTB-XL dataset.}
 \setlength{\tabcolsep}{0.4pt}
\begin{tabular}{lccccc}
\hline
 &
  Accuracy(\%) &
  \begin{tabular}[c]{@{}c@{}}ROC-AUC \\ Macro(\%)\end{tabular} &
  \begin{tabular}[c]{@{}c@{}}ROC-AUC \\ Samples(\%)\end{tabular} &
  \begin{tabular}[c]{@{}c@{}}Macro-F1 \\ score(\%)\end{tabular} &
  \begin{tabular}[c]{@{}c@{}}Samples-F1 \\ score(\%)\end{tabular} \\ \hline
Pałczynski et al.~\cite{palczynski2022study} & 66.20          & 84.40          & -               & 32.40          & -               \\
Smigiel et al.~\cite{smigiel2021deep}    & 67.60          & 86.10          & -               & 33.60          & -               \\
Feyisa et al.~\cite{feyisa2022lightweight}     & 96.20          & 92.00          & \textbf{-}      & 46.00          & -               \\
IMLENet               & 96.32          & 91.47          & \textbf{95.50} & 19.07          & 63.27          \\
GTN                    & 95.46          & 80.61          & 90.91          & 21.55          & 64.56          \\
Resnet101              & \textbf{96.87}          & 92.56          & 94.58          & 31.42          & 65.93          \\
ECGNet                 & 96.34          & \textbf{92.83}          & 94.14          & 33.53          & 65.90          \\
\textbf{MSW-TN}        & 96.78 & 92.65 & 94.40          & \textbf{47.57} & \textbf{68.27} \\ \hline
\end{tabular}
\label{tab2}
\end{table}

The results for Task 2, which involves classifying the subclass labels, are shown in Table~\ref{tab2}. The results for the first three models were obtained through our own experiments, while the results for the latter three models were directly cited from the literature. Our proposed model demonstrated superior performance in terms of Macro-F1 score and Samples-F1 score, achieving scores of 47.57\%, and 68.27\%, respectively. This is attributed to the fact that clinical diagnosis of subclasses such as Left Anterior/Left Posterior Fascicular Block (LAFB/LPFB) requires diagnostic criteria such as QRS complex widening~\cite{surawicz2009aha}, while the diagnosis of Left Atrial Overload/Enlargement (LAO/LAE) requires diagnostic criteria such as prolongation of the P-R interval~\cite{hancock2009aha}. These features necessitate the use of our large-scale window attention mechanism for feature extraction. In contrast, the diagnosis of Septal Hypertrophy (SEHYP), Anterior Myocardial Infarction (AMI), and Ischemic in Anterior Leads (ISCA) requires diagnostic criteria such as changes in the waveform of the ST segment~\cite{rautaharju2009aha}, which necessitates the use of our small-scale window attention mechanism for feature extraction. Our proposed MSW-Transformer can effectively extract these features from electrocardiogram signals and achieve better feature fusion.

\begin{table}[!ht]
\caption{\newline Model evaluation for classification of rhythm labels on the PTB-XL dataset.}
\setlength{\tabcolsep}{3pt}
\begin{tabular}{lccccc}
\hline
 &
  Accuracy(\%) &
  \begin{tabular}[c]{@{}c@{}}ROC-AUC \\ Macro(\%)\end{tabular} &
  \begin{tabular}[c]{@{}c@{}}ROC-AUC \\ Samples(\%)\end{tabular} &
  \begin{tabular}[c]{@{}c@{}}Macro-F1 \\ score(\%)\end{tabular} &
  \begin{tabular}[c]{@{}c@{}}Samples-F1 \\ score(\%)\end{tabular} \\ \hline
GTN       & 95.46          & 84.51          & 94.63          & 39.19          & 51.69          \\
Resnet101 & 95.15          & 84.40          & 97.68          & 41.89          & 88.72          \\
ECGNet    & 95.90          & \textbf{95.36} & 98.33 & 33.81          & 89.65          \\
IMLENet   & \textbf{98.49} & 94.97          & \textbf{98.61}          & 49.98          & 90.27          \\
\textbf{MSW-TN}        & 96.75          & 92.46          & 98.24          & \textbf{66.13} & \textbf{91.32} \\ \hline
\end{tabular}
\label{tab3}
\end{table}

For Task 3, Table~\ref{tab3} presents the classification results of rhythm labels obtained from our own experiments on the four compared models. Our proposed model outperforms the other models in terms of Macro-F1 score and Samples-F1 score, with scores of 66.13\% and 91.32\%, respectively. This is attributed to the diagnostic criteria of Atrial Fibrillation (AFIB) and Sinus Arrhythmia (SARRH), which require the extraction of features such as time differences between QRS complexes~\cite{rautaharju2009aha}. Our large-scale window is able to extract these features effectively. Similarly, the diagnosis of Sinus Tachycardia (STACH) and Atrial Flutter (AFLT) require the extraction of features such as changes in the ST segment~\cite{rautaharju2009aha}, which is achieved through our small-scale window. Our proposed MSW-Transformer is capable of extracting features from electrocardiogram signals efficiently, and performs effective feature fusion.

\begin{table}[ht]
\caption{\newline Model evaluation for classification of form labels on the PTB-XL dataset.}
\setlength{\tabcolsep}{2.9pt}
\begin{tabular}{lccccc}
\hline
 &
  Accuracy(\%) &
  \begin{tabular}[c]{@{}c@{}}ROC-AUC \\ Macro(\%)\end{tabular} &
  \begin{tabular}[c]{@{}c@{}}ROC-AUC \\ Samples(\%)\end{tabular} &
  \begin{tabular}[c]{@{}c@{}}Macro-F1 \\ score(\%)\end{tabular} &
  \begin{tabular}[c]{@{}c@{}}Samples-F1 \\ score(\%)\end{tabular} \\ \hline
GTN           & 92.99          & \textbf{93.00} & 84.12          & 10.84          & 24.31          \\
IMLENet       & 94.01          & 78.50          & \textbf{89.76} & 19.07          & 38.41          \\
Resnet101     & 91.96          & 64.04		  & 87.52 			& 23.16          & 36.89          \\
ECGNet         & \textbf{94.07} & 77.12          & 88.97          & 27.13          & 37.78          \\
\textbf{MSW-TN} & 92.34          & 78.32          & 87.31          & \textbf{34.60} & \textbf{50.07} \\ \hline
\end{tabular}
\label{tab4}
\end{table}

For Task 4, the classification results of form labels are presented in Table~\ref{tab4}, which were obtained through our own experiments on the four compared models. Our proposed model achieved the best performance in terms of Macro-F1 score and Samples-F1 score, with scores of 34.60\% and 50.07\%, respectively. This is because the diagnosis of Long QT-interval (LNGQT) in clinical practice requires the extraction of features such as the length of the QT interval~\cite{rautaharju2009aha}, and the diagnosis of Abnormal QRS (ABQRS) requires the extraction of features such as abnormal QRS intervals~\cite{rautaharju2009aha}, which can be effectively extracted using our large-scale window. Similarly, the diagnosis of Non-diagnostic T abnormalities (NDT), which requires the extraction of features such as inverted or low-amplitude T waves~\cite{rautaharju2009aha}, and ventricular premature complex, which requires the extraction of features such as ST-T wave changes~\cite{rautaharju2009aha}, can be extracted using our small-scale window. Our proposed MSW-Transformer can effectively extract features from electrocardiogram signals and achieve effective feature fusion.

\begin{table}[ht]
\caption{\newline Model evaluation for classification of all labels on the PTB-XL dataset.}
\setlength{\tabcolsep}{3pt}
\begin{tabular}{lccccc}
\hline
 &
  Accuracy(\%) &
  \begin{tabular}[c]{@{}c@{}}ROC-AUC \\ Macro(\%)\end{tabular} &
  \begin{tabular}[c]{@{}c@{}}ROC-AUC \\ Samples(\%)\end{tabular} &
  \begin{tabular}[c]{@{}c@{}}Macro-F1 \\ score(\%)\end{tabular} &
  \begin{tabular}[c]{@{}c@{}}Samples-F1 \\ score(\%)\end{tabular} \\ \hline
GTN          & 93.28          & 77.26          & 91.99          & 12.71          & 45.92          \\
Resnet101      & 96.19          & 85.06          & 93.18 			& 16.56          & 61.82          \\
ECGNet         & \textbf{97.86} & 86.33 		  & 95.21 			& 17.49          & 60.25          \\
IMLENet        & 97.85 			& 90.48          & \textbf{96.42} & 21.15          & 62.50          \\
\textbf{MSW-TN} & 96.30          & \textbf{92.35} & 95.05          & \textbf{34.29} & \textbf{63.19} \\ \hline
\end{tabular}
\label{tab5}
\end{table}

For Task 5, the classification results of all labels are presented in Table~\ref{tab5}, which were obtained through our own experiments on the four compared models. Our proposed model achieved the best performance in terms of Macro ROC-AUC, Macro-F1 score, and Samples-F1 score, with scores of 92.35\%, 34.29\%, and 63.19\%, respectively. The results of the all labels classification demonstrate that our model is capable of effectively classifying multiple categories and can handle features of different scales, while maintaining high efficiency.

\subsection{Ablation Experiment}

\begin{figure}[!ht]
\centering 
\centerline{\includegraphics[width=1.2\linewidth]{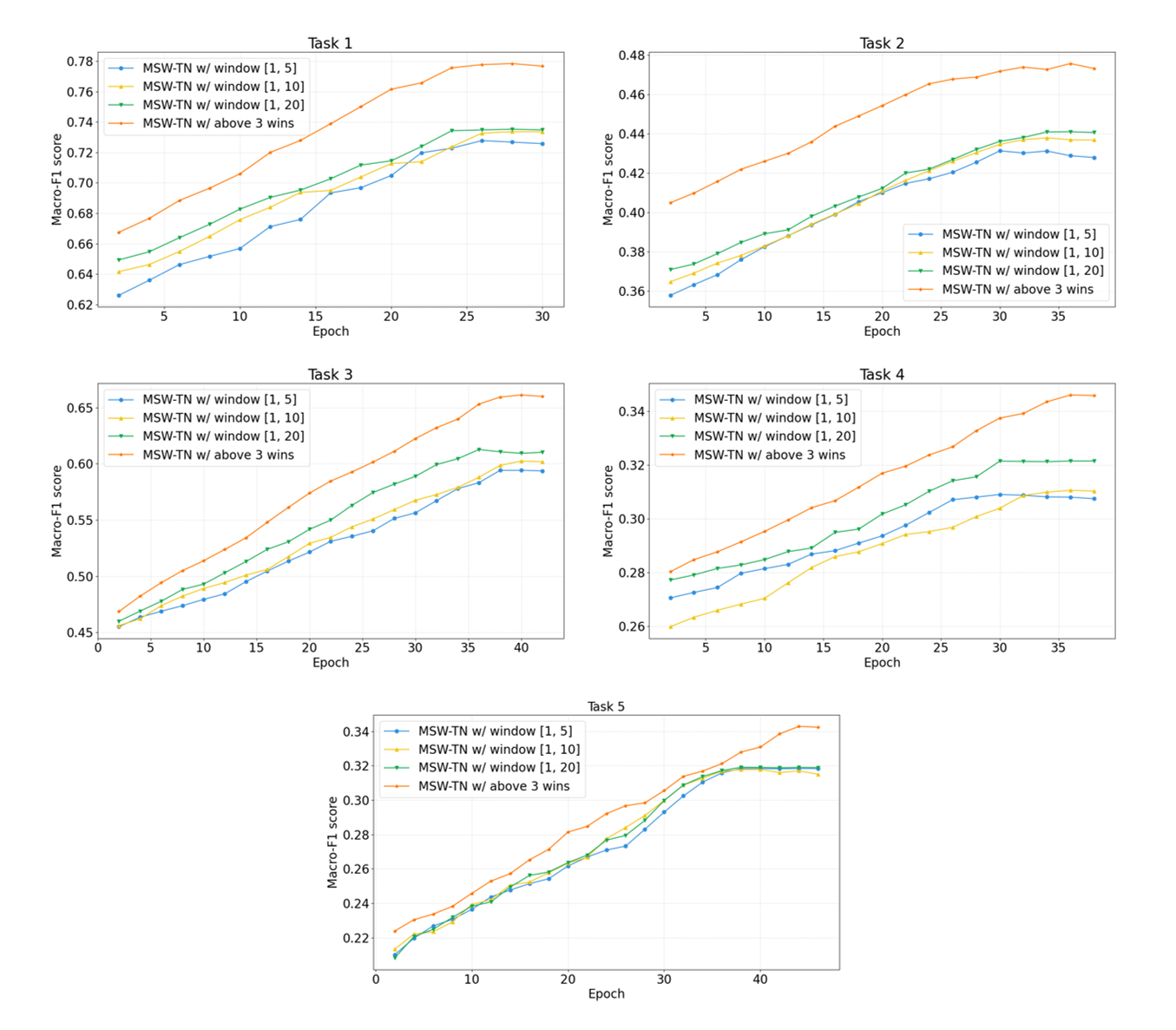}}
\caption{Comparison of Macro-F1 score for classification tasks using different scales of window shift attention mechanisms (including single-window shift attention mechanisms with window sizes of [1, 5], [1, 10], and [1, 20], and the shift attention mechanism with three windows).}
\label{fig9}
\end{figure}

\begin{figure}[!ht]
\centering 
\centerline{\includegraphics[width=1.2\linewidth]{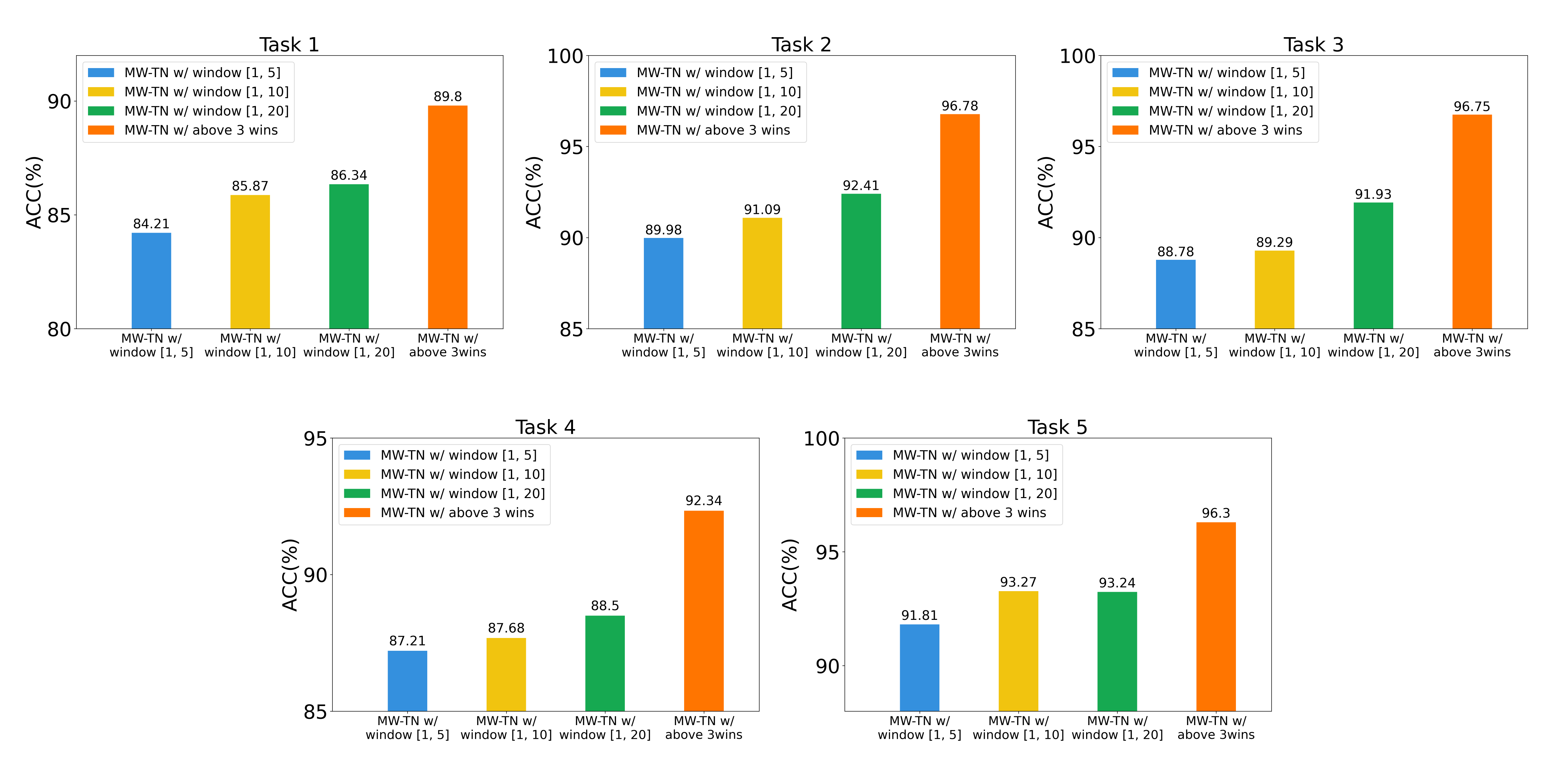}}
\caption{ Comparison of Accuracy for classification tasks using different scales of window shift attention mechanisms (including single-window shift attention mechanisms with window sizes of [1, 5], [1, 10], and [1, 20], and the shift attention mechanism with three windows).}
\label{fig10}
\end{figure}

To evaluate the superiority of the MSW-Transformer, we conducted three experiments using the PTB-XL dataset and replicated the same five classification tasks as in our prior studies. We substituted the multi-window shift attention mechanism with three single-window shift attention mechanisms and obtained the classification results for Macro-F1 score and accuracy, which are displayed in Fig.~\ref{fig9} and Fig.~\ref{fig10}. Our findings revealed that the single-scale window shift attention mechanisms, with window sizes of [1, 5], [1, 10], and [1, 20], were inadequate for achieving reliable classification of electrocardiogram signals. However, our proposed multi-scale window shift attention mechanism significantly enhanced the classification performance. Specifically, the classification results for all five tasks using this mechanism attained the highest Macro-F1 score and accuracy, thereby validating the effectiveness of the MSW-Block. By utilizing multiple windows of diverse scales, it can efficiently capture features of various scales and boost feature information through complementary data, improving in classification performance.

\begin{figure}[!ht]
\centering 
\centerline{\includegraphics[width=1.2\linewidth]{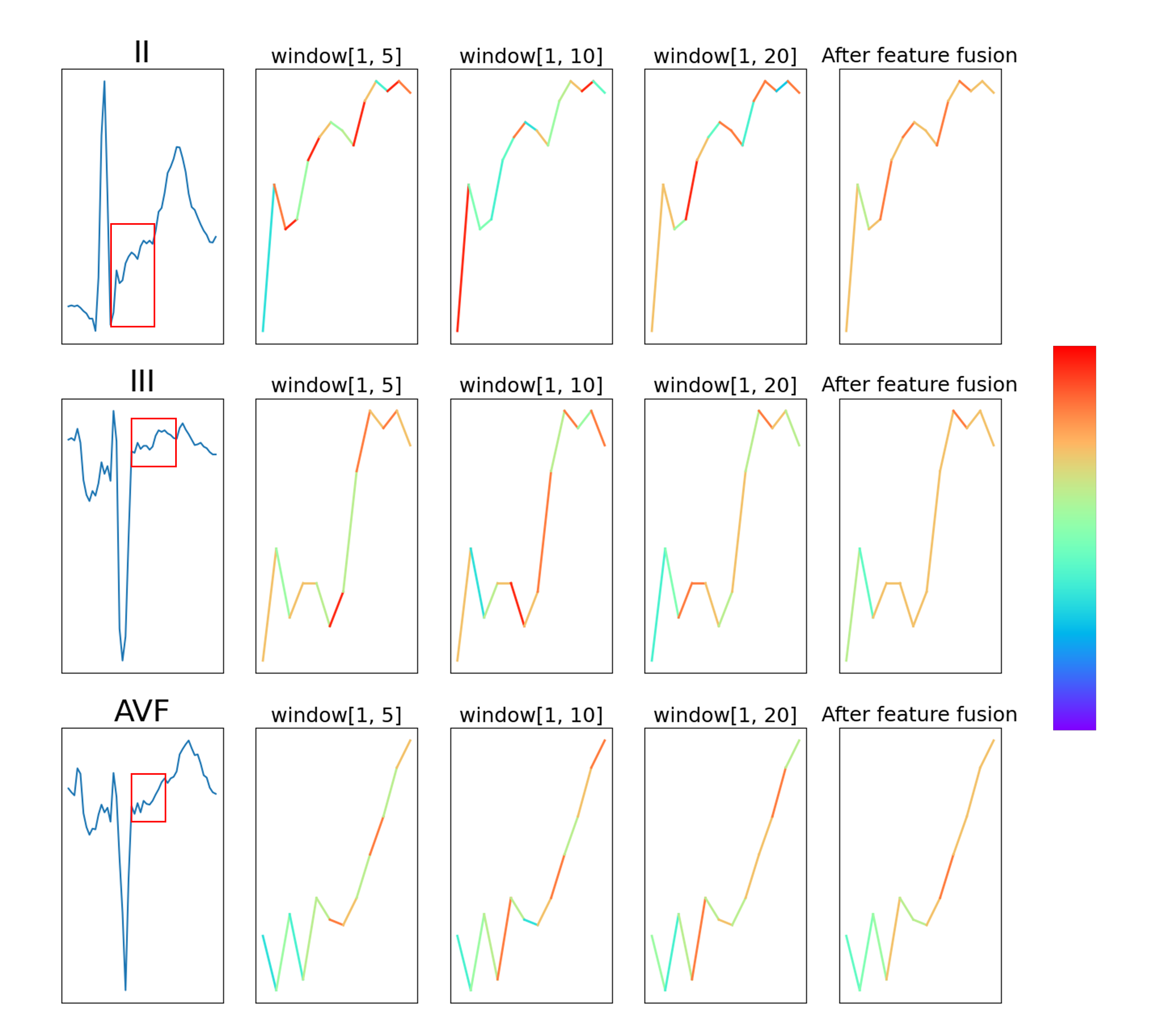}}
\caption{Visualization of standardized attention scores for a partial electrocardiogram (ECG) of a patient with inferior myocardial infarction (IMI) (the red box indicates the ST segment) (red indicates a higher attention score, and blue indicates a lower attention score).}
\label{fig11}
\end{figure}

In order to further demonstrate the limitations of feature extraction using a single window and the feasibility and superiority of MSW-Feature fusion, we visualize the attention scores of three individual windows and the fused feature attention score. Taking patients diagnosed with inferior myocardial infarction (IMI) as an example, in clinical practice, the diagnosis of IMI is based on ST segment elevation in leads II, III, and aVF~\cite{morris2002abc,edhouse2002acute}. As shown in Fig.~\ref{fig11}, we extracted the segments containing the ST segment in leads II, III, and aVF of the patient (the red box indicates the ST segment) and compared the attention scores of the three individual windows and the fused feature attention score (red indicates the highest attention score, and blue indicates the lowest attention score). It can be observed that there are some segments with low attention scores in the ST segment when using a single window, indicating that these features have not been fully attended to. However, after feature fusion, the attention scores of the entire ST segment are consistently high. This suggests that feature extraction using a single window is incomplete, while feature fusion can improve the accuracy and comprehensiveness of feature extraction, enabling the model to classify more accurately.

\section{DISCUSSION}

\subsection{Applicability of the Model}

The ECG signal is a quasi-periodic voltage signal ~\cite{nagendra2011application}, and clinical classification of ECGs is also usually based on certain characteristics of a particular heartbeat cycle of the signal. For example, ECG judgments in patients with left posterior branch block (LAFB) are based on having small R waves and deep S waves in leads II, III, and aVF, small Q waves and high R waves in leads I and aVL \cite{GOLDBERGER200673}, and ECG judgments in patients with left ventricular hypertrophy (LVH) are based on the typical I, aVL, and V4-V6 leads of increased R-wave amplitude and increased S-wave depth in leads III, aVR, and V1-V3 ~\cite{morrison2007evaluation}. Therefore, it is well suited for feature extraction within a window attention mechanism in ECG signals.

Considering the complexity of ECG signals, it is often difficult to determine an optimal window size for feature extraction. Thus, we propose the MSW-Block as the optimal choice. This approach extracts features of different scales within windows of varying sizes, which can potentially enhance the performance and generalization ability of the model. By incorporating a multi-window attention mechanism, the MSW-TN can effectively handle the highly complex periodicity of ECG signals and extract richer feature information, leading to enhanced accuracy and reliability in ECG classification and related tasks. Therefore, our proposed MSW-TN has significant potential for practical applications in addressing ECG signal classification and other related problems.

\subsection{Model Complexity}

We conducted experiments to compare the global attention mechanism and the multi-window attention mechanism in terms of model parameter size, theoretical computational complexity, and inference speed. We found that using the multi-window attention mechanism can effectively reduce the model size and complexity, and speed up the inference process, which is one of the advantages of our proposed model. Furthermore, we found that the global attention mechanism is not suitable for ECG signal classification scenarios. On the contrary, for small-scale data tasks such as electrocardiogram signal classification, traditional Transformer models are prone to overfitting, which explains why they can achieve good results in large-scale datasets in the computer vision field. However, ECG signal datasets are limited, so we need a more suitable model to perform ECG classification.

The traditional Transformer model is characterized by its large number of layers, parameters, and computational complexity, which often results in slow inference speeds in many real-world scenarios, thereby limiting its practical applicability. To address these challenges, we propose the MSW-Transformer model. This model reduces the attention extraction operations from multiple layers to a single layer, resulting in a significant reduction in model size and complexity, faster inference speeds, and prevention of overfitting. Our experimental results demonstrate that this approach not only performs well in terms of model performance but also provides a substantial theoretical basis for the clinical application of electrocardiogram signal recognition. Thus, this approach is of significant practical significance and has promising prospects for widespread application in real-world scenarios.

\subsection{MSW-Feature fusion}
There are two basic methods for feature fusion: direct feature concatenation and combining feature vectors into composite vectors. In this study, we combine both methods and perform learnable fusion of feature vectors obtained from each window's attention mechanism. This selection process helps the model better classify electrocardiogram signals.

Drawing from the aforementioned considerations, MSW-Feature fusion executes a learnable fusion of feature vectors obtained from each window to effectively combine multiple features and enhance the representation of ECG signal features. This process is learnable, and the optimal feature combination is selected during the iterative process, thereby improving the accuracy of classification. Experimental results corroborate the efficacy of our proposed MSW transformer model in ECG signal classification tasks. Compared to other methods, our approach yields higher Macro-F1 scores and Samples-F1 scores. These findings underscore the effectiveness and feasibility of our proposed feature fusion method and emphasize the significance of this fusion strategy for multi-task classification.

\subsection{Effectiveness of the Model}
\begin{figure}[!ht]
\centering 
\centerline{\includegraphics[width=0.8\linewidth]{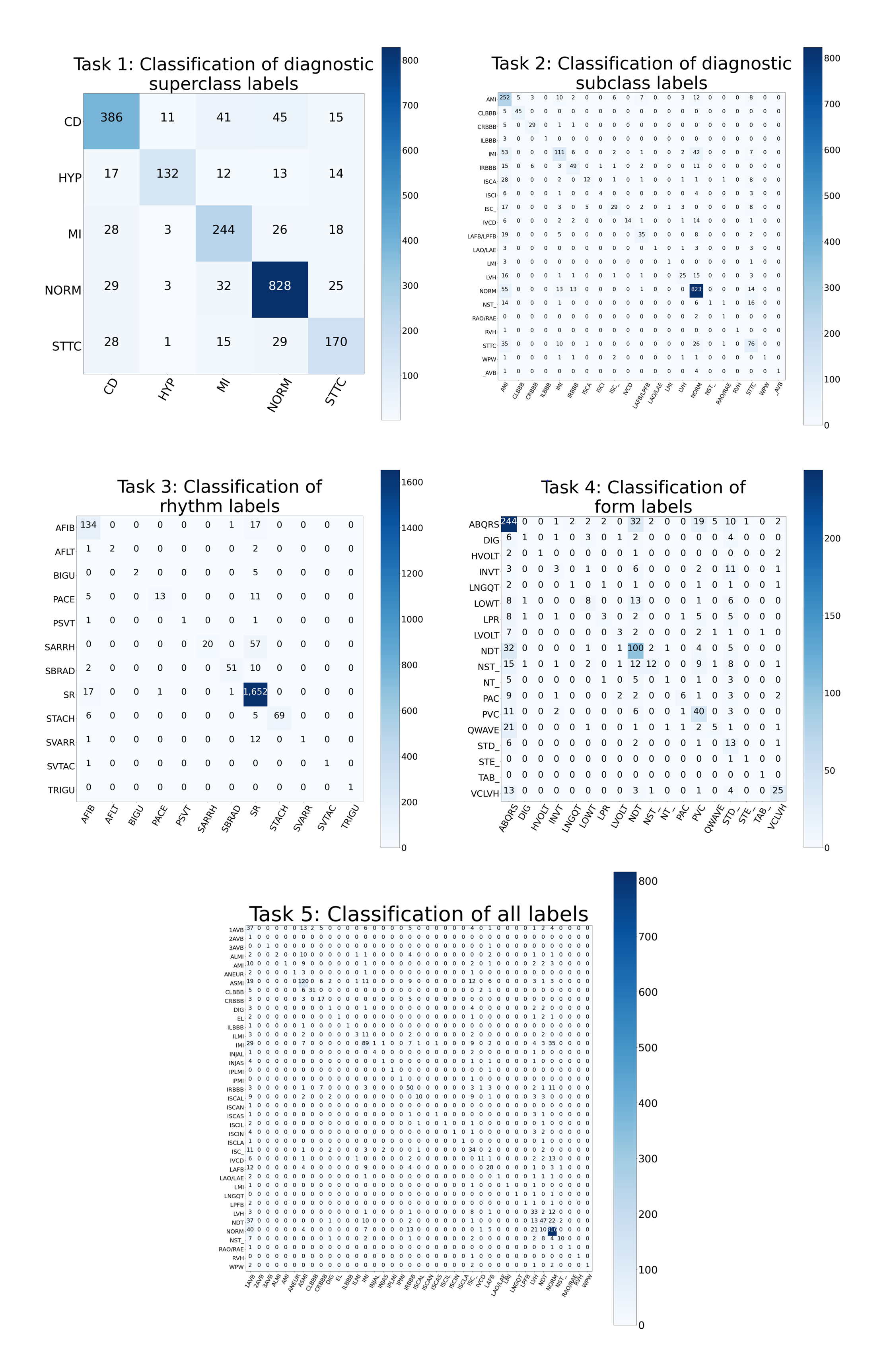}}
\caption{Confusion matrix for classification tasks 1-5.}
\label{fig12}
\end{figure}

Our proposed MSW-TN achieves the best Macro-F1 score and Samples-F1 score in all five classification tasks, as shown in Fig.~\ref{fig12}, which presents the confusion matrix results of our five classification tasks and demonstrates the effectiveness of MSW-TN. Specifically, based on the calculations from Fig.~\ref{fig12}, the Macro-F1 scores of the Norm class in classification task 1, the CLBBB class in classification task 2, and the SR class and STACH class in classification task 3 are 89.13\%, 90.00\%, 95.96\%, and 92.62\%, respectively. This indicates that the classifier has a low misclassification rate and performs well in classification tasks, with high reliability and accurate predictions for practical applications.

However, due to the data imbalance in the PTB-XL dataset, the model's learning on certain labels is insufficient, leading to a bottleneck in classification performance. For example, as shown in Fig.~\ref{fig12}, the small number of sub-labels such as ISCA, NST, LAFB/LPFB makes it difficult for the model to learn these labels. In the identification of ISCA (ischemia in the front lead), 28 cases were identified as AMI (acute myocardial infarction), which may be due to similarities between the diagnostic criteria for these two diseases in ECG signals, such as ST segment elevation in leads V1-V4. In the identification of NST (non-specific ST/T wave changes), 16 cases of NST were identified as STTC (ST/T wave changes due to myocardial ischemia), which may be due to similarities in the diagnosis of these two diseases in ECG signals, such as ST segment depression or elevation and T wave inversion in most leads~\cite{morris2002abc}. In the identification of LAFB/LPFB (left anterior fascicular block/left posterior fascicular block), 19 cases were identified as AMI, which may be due to similarities in the diagnostic criteria for these two diseases in ECG signals, such as Q wave prolongation and T wave inversion in leads V1, V5, and V6~\cite{morris2002abc}. As some diseases have a small amount of data and similar reasons for differentiation in clinical diagnosis, our model faces greater difficulty in classifying them. This indicates that our approach needs to further improve the accuracy of classification for diseases with similar features.

\section{Conclusion}

In this study, we propose a novel classification model, the MSW-Transformer network, for multi-channel ECG signals. Our main contributions are: (1) the proposition of a transformer block based on a sliding shift attention mechanism using multiple window scales, (2) the introduction of a feature fusion method, the MSW-Feature fusion, and (3) the visualization of the attention scores of different windows in the model. We compare the attention scores of the model with the rules for myocardial infarction diagnosis by cardiac experts, which further demonstrates the interpretability and reliability of the model and aids in understanding its clinical decision-making applications. We evaluate our model in a five-class classification task on the PTB-XL dataset, and our results show that our model outperforms several existing models. Notably, our algorithm has lower model parameters and complexity compared to traditional transformer models, with faster inference speed. This study provides a robust foundation for developing extensive cardiac assessment models and has practical application value.



\bibliographystyle{elsarticle-num} 
\bibliography{elsarticle-template-num}





\end{document}